\documentclass[aps,prl,twocolumn]{revtex4-1}
\pdfoutput=1
\usepackage[T1]{fontenc}
\usepackage[english]{babel}
\usepackage{float}
\usepackage[utf8]{inputenc}
\usepackage{pslatex}   
\usepackage{amsmath}

\makeatletter

\newcommand{\Felec}{F_\mathrm{elec}}

\newcommand{\Vbias}{V_\mathrm{bias}}
\newcommand{\Vcpd}{V_\mathrm{cpd}}

\newcommand{\dcdz}{\frac{\partial C}{\partial z}}

\newcommand{\Vdc}{V_\mathrm{dc}}
\newcommand{\Vac}{V_\mathrm{ac}}

\newcommand{\df}{\Delta f}

\newcommand{\Fin}{F_\mathrm{in}}
\newcommand{\Fout}{F_\mathrm{quad}}
\newcommand{\wm}{\omega_\mathrm{m}}
\newcommand{\Fwm}{F_\mathrm{\wm}}
\newcommand{\wel}{\omega_\mathrm{el}}
\newcommand{\fm}{f_\mathrm{m}}
\usepackage{babel}
\usepackage{graphicx}

\graphicspath{{./Figures}{}}

\makeatother
\begin{document}

\date{\today}

\title{Force-gradient sensitive Kelvin probe force microscopy 
by dissipative electrostatic force modulation}

\author{Yoichi Miyahara}
\email[Corresponding author:]{yoichi.miyahara@mcgill.ca}
\author{Peter Grutter}
\affiliation{Department of Physics,
  McGill University, 3600 rue University, Montreal, Quebec, Canada H3A 2T8}

\begin{abstract}
We report a Kelvin probe force microscopy (KPFM) implementation 
using the dissipation signal of a frequency modulation atomic
force microscopy that is capable of detecting 
the gradient of electrostatic force rather than electrostatic force. 
It features a simple implementation and faster scanning
as it requires no low frequency modulation.
We show that applying a coherent ac voltage with two times the cantilever 
oscillation frequency induces the dissipation signal proportional to 
the electrostatic force gradient 
which depends on the effective dc bias voltage including
the contact potential difference.
We demonstrate the KPFM images of a MoS$_2$ flake taken 
with the present method is in quantitative agreement with
that taken with the frequency modulated Kelvin probe force microscopy technique.
\end{abstract}

\maketitle

We recently reported a Kelvin probe force microscopy (KPFM) implementation 
(D-KPFM) in which the dissipation signal 
of a frequency modulation atomic force microscopy (FM-AFM)
is used for dc-bias voltage feedback \cite{Miyahara2015}.
In KPFM, a contact potential difference (CPD) between the tip and sample, $\Vcpd$, 
is measured by finding a dc bias voltage, $\Vbias$, 
nullifying a capacitive electrostatic force, $\Felec$, 
that is proportional to the effective dc potential difference, 
$\Vdc=(\Vbias - \Vcpd)$.
  In order to detect $\Felec$ in the presence of other force 
components such as van der Waals force and chemical bonding force, 
$\Felec$ is usually modulated by superposing an ac bias voltage
to $\Vbias$
and the resulting modulated component of the measured force 
(or force gradient) is detected by lock-in detection \cite{Nonnenmacher91}.
While the modulated frequency shift signal is used in the case of 
the conventional KPFM with FM-AFM (FM-KPFM) \cite{Kitamura2000}, 
in the case of D-KPFM, $\Felec$ is detected through the dissipation signal of FM-AFM,
which is the amplitude of cantilever excitation signal.
The dissipation is induced by applying a coherent sinusoidal ac voltage 
with the frequency of the tip oscillation 
which is $90^\circ$ out of phase with respect to the tip oscillation.
The induced dissipation signal can be used for 
KPFM voltage feedback loop 
as it is proportional to the effective dc bias voltage, $\Vdc=(\Vbias -\Vcpd)$.

Here we report another D-KPFM implementation ($2\omega$D-KPFM)
where the induced dissipation is proportional to the electrostatic force gradient,
resulting in the identical potential image contrast as is obtained by FM-KPFM.
In $2\omega$D-KPFM, the frequency of the applied ac voltage is 
two times that of tip oscillation.

The electrostatic force between two conductors connected 
to an ac and dc voltage source, $\Felec$,
is described as follows \cite{Kantorovich00}:
\begin{eqnarray} 
       \Felec &=& \frac{1}{2}\dcdz \{\Vbias -\Vcpd +\Vac \cos (\wel t + \phi)\}^2 \nonumber\\ 
       & = & \alpha\{\Vdc + \Vac\cos (\wel t + \phi)\}^2 \label{eq:Felec}
 \end{eqnarray}
where $C$ is the tip-sample capacitance 
and $\alpha \equiv \frac{1}{2}\dcdz$.  
$\Vac$ and $\wel$  are the amplitude, angular frequency of the ac bias voltage 
and $\phi$ is the phase of the ac bias voltage with respect to the tip oscillation.
$z$ is the position measured from the sample surface 
and is oscillating around its mean position, 
$z_0$, expressed as $z(t)= z_0 + A\cos(\wm t)$ 
with $\wm$ and $A$ being its oscillation angular frequency and amplitude,
respectively.
We assume that the tip oscillation is driven by another means 
such as piezoacoustic or photothermal excitation \cite{Labuda2012a}
and its frequency, $\fm=\wm/2\pi$, 
is chosen to be the fundamental flexural resonance frequency, $f_0$.
In FM-AFM, 
the resonance frequency shift, $\Delta f$, and dissipation signal, $g$,
are proportional to the in-phase, $\Fin$, and quadrature, $\Fout$, components 
of the fundamental harmonic component (in this case $\wm$) of the oscillating 
force, respectively \cite{Holscher2001,Kantorovich2004, Sader2005}
such as $\Delta f=-\frac{1}{2} \frac{f_0}{kA} \Fin$ and $g=g_0 (1-\frac{Q}{kA}\Fout)$where $k$ and $Q$ are the spring constant and quality factor of the cantilever, 
and $g_0$ is the dissipation signal determined 
by the intrinsic dissipation of the cantilever.
As we have already pointed out \cite{Miyahara2015}, 
the distance dependence of $\alpha (z) \equiv \frac{1}{2}\dcdz(z)$ 
has to be taken into account 
in order to obtain the correct forms of $\Fin$ and $\Fout$
acting the oscillating tip that is subject to 
the coherent oscillating electrostatic force.
It is the interaction between the mechanical oscillation of the tip and the coherent 
oscillating electrostatic force that leads to the following non-obvious results
\cite{Sugawara2012}.

Here, we consider the case of $\wel = 2\wm$ which has previously been suggested 
by Nomura {\it et al.} \cite{Nomura2007} for a similar 
but different KPFM implementation with dissipative electrostatic force modulation
\cite{Fukuma2004}.
\begin{figure}[t]
  \centering
  \includegraphics[width=80mm]{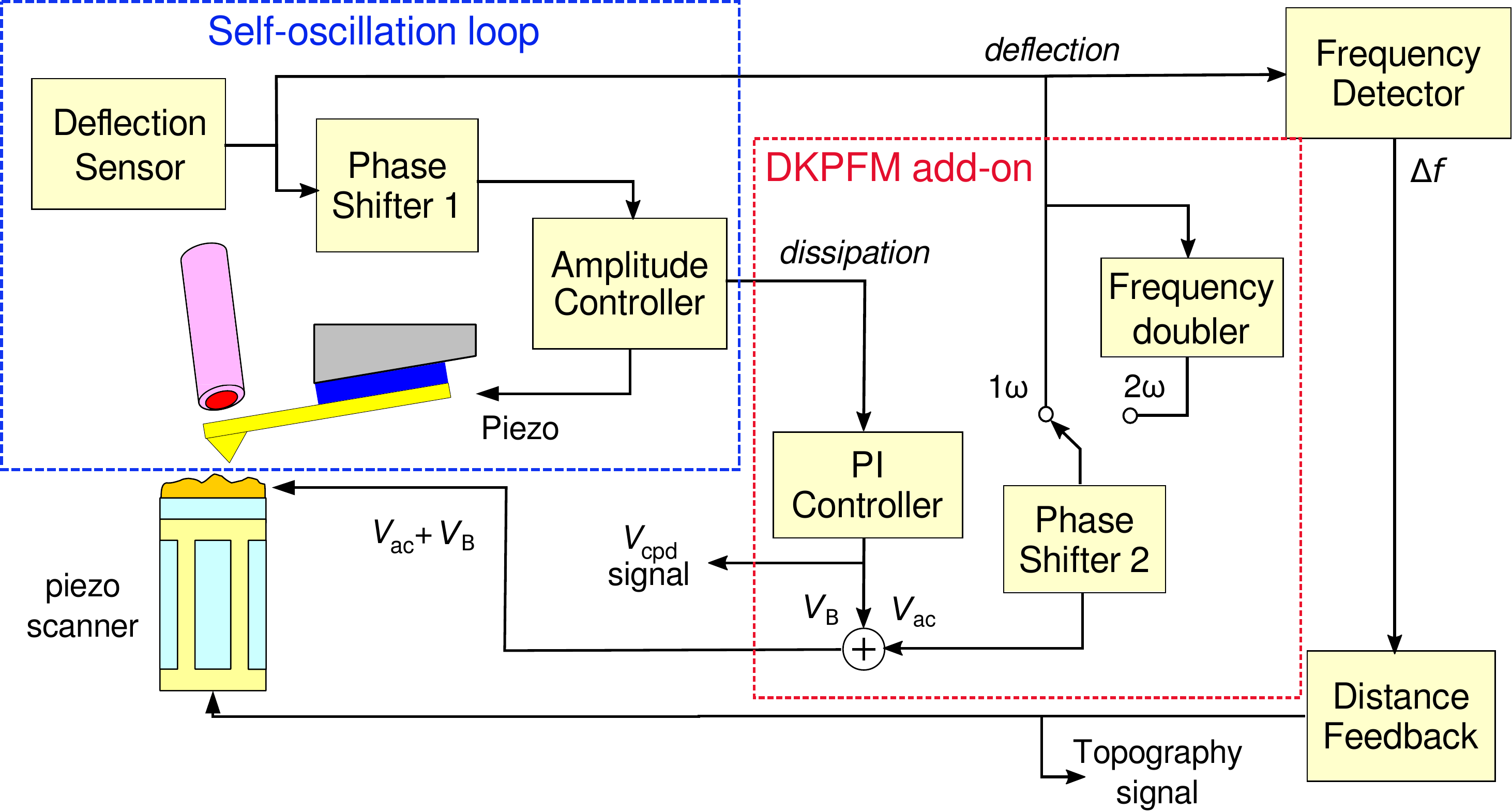}
  \caption{Block diagram of the experimental setup for force-gradient sensitive 
D-KPFM measurements.
    \label{fig:DKPFMDiagram2}}
\end{figure}
By expanding $\alpha(z)$ around the mean tip position, $z_0$, 
and taking the first order term,
$\alpha$ is expressed as 
{$  \alpha (z) \approx \alpha(z_0) + \alpha ' (z-z_0) = \alpha_0 + \alpha ' A \cos(\wm t)
  $\label{eq:alpha}}.
Substituting this into Eq.~\ref{eq:Felec} with setting $\wel = 2\wm$ 
and gathering all the terms with $\wm$ components yields the following result
(detailed derivation available in Supplemental Information):
\begin{eqnarray*}
  \Fwm(t) & = &   \alpha' A \left\{ \left(\Vdc ^2 + \frac{\Vac^2}{2}\right) \cos (\wm t)
      +  \Vdc \Vac  \cos (\wm t + \phi)\right\}\\
       & = & \alpha' A \left\{\Vdc ^2 +  \cos(\phi) \Vdc\Vac +\frac{\Vac^2}{2} \right\}\cos(\wm t) \\ 
  & &    -\alpha' A \Vdc \Vac  \sin(\phi) \sin(\wm t)\\
    & = & \Fin \cos (\wm t) + \Fout \sin(\wm t)
\end{eqnarray*}
where
\begin{equation}
  \Fin =  \alpha'A \left(\Vdc + \frac{\Vac}{2} \cos\phi   \right)^2
  + \frac{\alpha'A }{2}\left(1 - \frac{\cos^2\phi}{2} \right) \Vac^2
  \label{eq:Fin}
\end{equation}
\begin{equation}
  \Fout = - \alpha'A  \Vdc\Vac  \sin\phi 
    \label{eq:Fquad}
\end{equation}
We notice that $\Fout$ is proportional to $\alpha '$ rather than $\alpha$,
indicating $\Fout$ is now sensitive to the electrostatic force gradient.
It is interesting to notice that 
the apparent shift of $\Vcpd$ 
is just determined by $\phi$ and $\Vac$ and 
does not depend on $\alpha$ 
in contrast to $1\omega$D-KPFM (See Eqs.~11-13 in Ref.~\cite{Miyahara2015}).
A key aspect of this new technique is the absence of such a shift 
in $\Fout$ for any $\phi$,
which is advantageous 
as it means that the accurate setting of $\phi$ is not required 
for accurate CPD measurements.
$\phi$ just affects the magnitude of $\Fout$,
which determines the signal-to-noise ratio of the CPD measurements.
When $\phi =90^\circ$, we obtain
\begin{equation*}
  \Fin =  \alpha'A \left(\Vdc^2 + \frac{\Vac^2}{2} \right) 
  = \alpha'A (\Vbias - \Vcpd)^2 +\frac{\alpha'A }{2}\Vac^2
\end{equation*}
\begin{equation*}
  \Fout = - \alpha'A  \Vdc\Vac =- \alpha'A  (\Vbias - \Vcpd)\Vac
\end{equation*}
In this case, the apparent shift of CPD in $\Fin$ vanishes 
and the slope of $\Fout$-$(\Vbias - \Vcpd)$ curve takes its maximum value.
Notice that $\Fout$ behaves exactly in the same way as the lock-in demodulated
frequency shift signal used in FM-KPFM.
When  $\Vbias = \Vcpd$, 
the dissipation signal turns back to its original value, $g_0$.
It is therefore possible to use the dissipation signal, $g$, 
for the KPFM bias voltage feedback 
by choosing $g_0$ as its control setpoint value.

\begin{figure}[t]
  \centering
  \includegraphics[width=78mm]{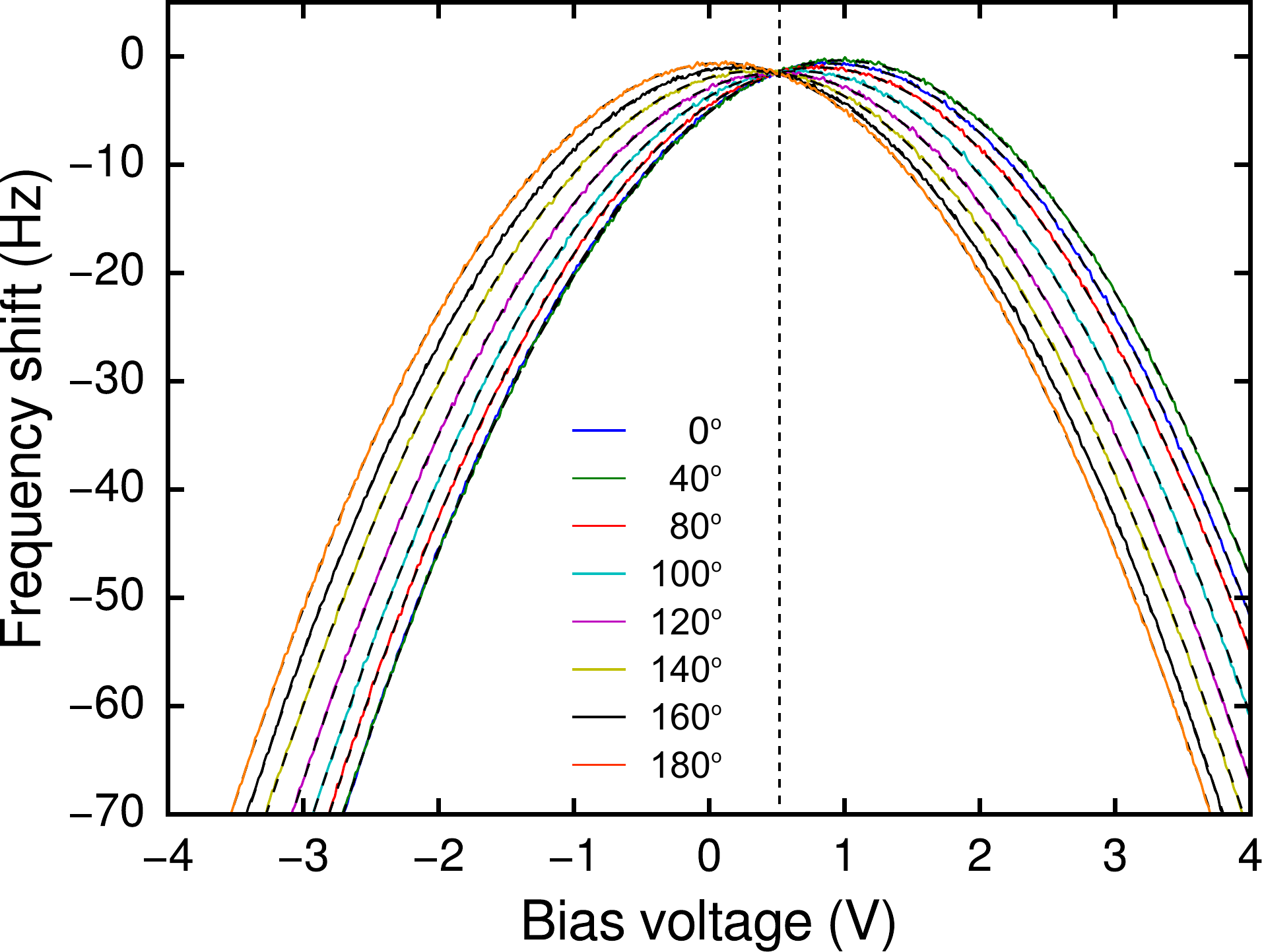}\\
  \vspace*{3mm}
  \includegraphics[width=78mm]{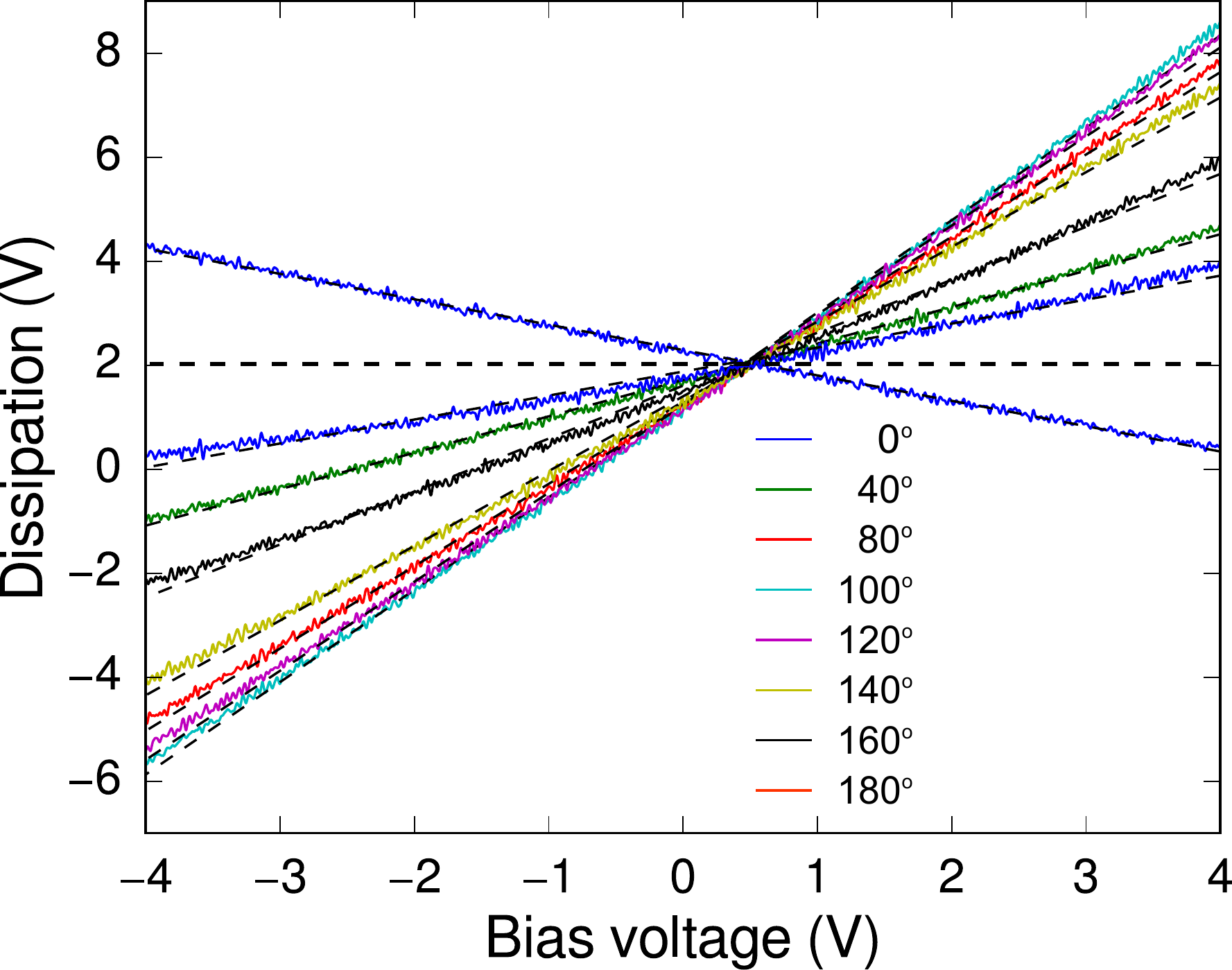}   
  \caption{(a) Frequency shift, $\df$, and (b) dissipation signal, $g$, 
  versus dc bias voltage, $\Vbias$, curves 
  taken with a coherent sinusoidally oscillating voltage 
  with $\wel=2\wm$, $\Vac=1$~V 
  and various $\phi$,
  applied to a template stripped gold substrate.
  The vertical dashed line in (a) indicates $\Vcpd$ which is measured 
  as the voltage coordinate of the parabola vertex without $\Vac$.
  The horizontal dashed line in (b) indicates the dissipation without $\Vac$.
  In both figures, each of dashed lines represent fitted curves 
  assuming a parabola for $\df$ and a linear line for $g$ as indicated
  in Eq.~\ref{eq:Fin} and \ref{eq:Fquad}, respectively.
  The oscillation amplitude of the tip was 10~nm$_\text{p-p}$ 
  and the quality factor of the cantilever was 25000.
  \label{fig:df-V}}
\end{figure}

Figure~\ref{fig:DKPFMDiagram2} depicts the block diagram of the experimental 
setup used for both $1\omega$ and $2\omega$D-KPFM measurements.
The setup is based on the self-oscillation mode FM-AFM system \cite{Albrecht1991}
and three additional components, a frequency doubler, a phase shifter,
a proportional-integrator (PI) controller, are required for D-KPFM operation.
The amplitude controller composed of a root-mean-square (RMS) 
amplitude detector and
a PI controller is used to keep an oscillation amplitude constant.
 The output of the amplitude controller is the dissipation signal
which is used for controlling $\Vbias$.
The detection bandwidth of the RMS amplitude detector 
is extended to about 3~kHz.
 In order to produce a sinusoidal ac voltage with two times the tip oscillation frequency,
the sinusoidal deflection signal from the cantilever deflection sensor 
is first fed into a frequency doubler.
Then the output of the frequency doubler passes through 
the additional phase shifter (Phase Shifter 2), 
which serves to adjust the phase of the ac voltage, $\phi$.
The dissipation signal acts as the input signal to the PI controller,
which adjusts $\Vbias$ 
to maintain a constant dissipation equal to the value without $\Vac$ applied,
$g_0$ \cite{note:HF2LI}.

We used a JEOL JSPM-5200 atomic force microscope for the experiments 
with the modifications described in Ref.~\cite{Miyahara2015}.
An open source scanning probe microscopy control software GXSM
was used for the control and data acquisition \cite{Zahl2010}.
A commercial silicon AFM cantilever (NSC15, MikroMasch) 
with a typical spring constant of about 28~N/m and resonance frequency 
of $\sim 300$~kHz was used in high-vacuum environment with
the pressure of $1 \times 10^{-7}$~mbar.

In order to validate Eqs.~\ref{eq:Fin} and \ref{eq:Fquad},
$\df$-$\Vbias$ and $g$-$\Vbias$ curves were measured 
while a coherent sinusoidally oscillating voltage 
with $\wel = 2\wm$, $\Vac=1$~V (2~V$_\text{p-p})$
and various phases, $\phi$
is superposed with $\Vbias$.
\begin{figure}[]
  \includegraphics[width=82mm]{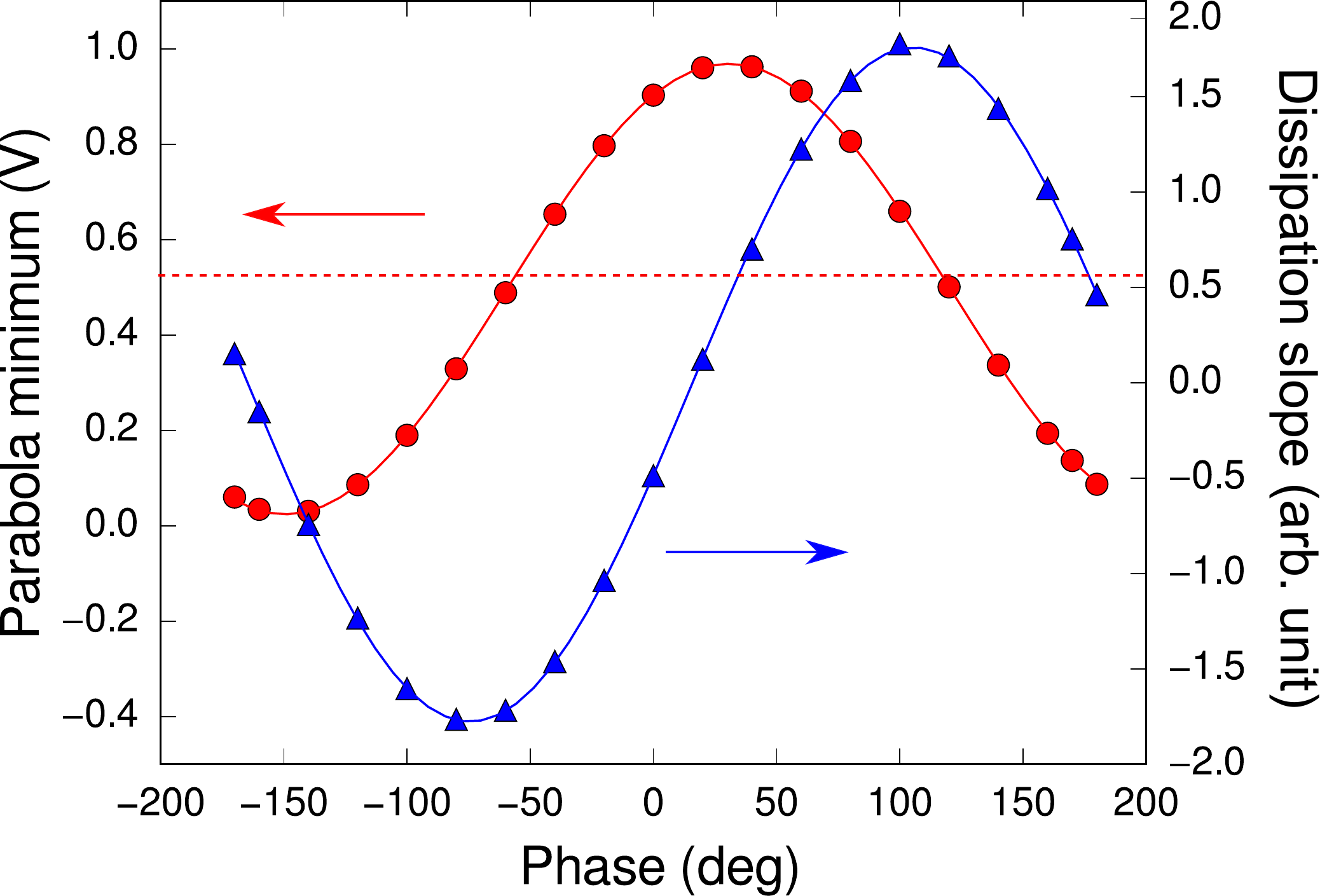}   
  \caption{Voltage coordinate of the vertices of the measured $\df$-$\Vbias$ curves (parabola minimum) 
  (red circles) extracted from the results Fig.~\ref{fig:df-V}(a)
  and the slope of dissipation-$\Vbias$ curves (blue triangles) extracted from
  Fig.~\ref{fig:df-V}(b). 
  Each solid line represents the fitted curve with the cosine function (Eq.~\ref{eq:Fin})
  for the parabola minimum and with the sine function (Eq.~\ref{eq:Fquad})
  for the dissipation slope.
  The horizontal dashed line indicates the voltage coordinate of the parabola 
  measured without the ac bias voltage.
  \label{fig:fit_parameters}}
\end{figure}
Figure~\ref{fig:df-V}(a) and (b) show simultaneously measured 
$\df$ and $g$ versus $\Vbias$ curves, respectively.
The curves are taken on a template stripped gold surface.
A fitted curve with a parabola for each of the $\df$-$\Vbias$ curves
(Eq.~\ref{eq:Fin})
or with a linear line for each of the $g$-$\Vbias$ curves (Eq.~\ref{eq:Fquad}) 
is overlaid on each experimental curve,
indicating a very good agreement between the theory and experiments.
As can be seen in Fig.~\ref{fig:df-V}(a) and (b), 
the position of the parabola vertex shifts 
both in $\Vbias$ and $\df$ axes
and the slope of $g$-$\Vbias$ curve changes 
systematically with the varied phase, $\phi$. 
 For further validating the theory,
the voltage coordinate of the parabola vertex (parabola minimum voltage) 
of each $\df$-$\Vbias$ curve and 
the slope of each $g$-$\Vbias$ curve are plotted against $\phi$
in Fig.~\ref{fig:fit_parameters}.
Each plot is overlaid with a fitted curve (solid curve) with the cosine function
(see Eq.~\ref{eq:Fin})
for the parabola minimum voltage 
and with the sine function (Eq.~\ref{eq:Fquad}) 
for the dissipation slope,
demonstrating an excellent agreement between the experiment and theory.
The $\Vbias$ dependence of the frequency shift coordinate 
of the parabola vertices 
(frequency shift offset) also shows a very good agreement with the theory 
(second term of Eq.~\ref{eq:Fin})
(The fitting result is available in Supplemental Information).
The parabola minimum voltage versus phase curve intersects 
that of $\df$-$\Vbias$ without ac bias voltage at $\phi = 121^\circ$.
The deviation from the theoretically predicted value of 90$^\circ$ 
is due to the phase delay in the detection electronics.
We also notice that the amplitude of parabola minimum versus phase curve is
0.472~V which is in good agreement with 0.5~V predicted by the theory
($\Vac /2$ in Eq.~\ref{eq:Fin}).

\begin{figure*}[]
  \centering
  \includegraphics[width=170mm]{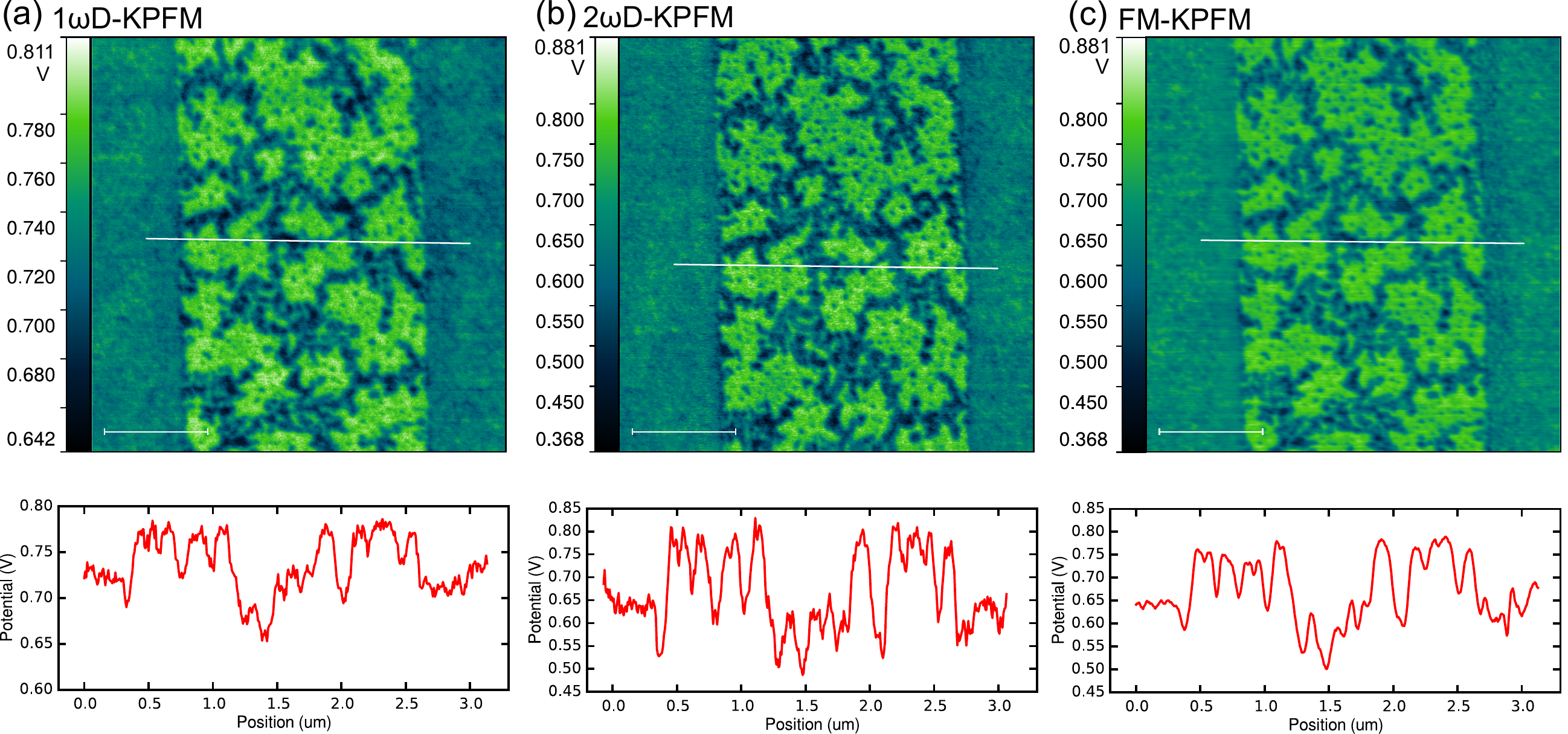}
  \caption{Potential images of patterned Mo$S_2$ on SiO$_2$/Si substrate 
        by (a) $1\omega$D-KPFM  ($\df = -10.3$~Hz, $A=5$~nm$_\text{p-p}$, 
        $\Vac=50$~mV),
        (b) $2\omega$D-KPFM  ($\df = -8.3$~Hz, $A=5$~nm$_\text{p-p}$, 
        $\Vac=1$~V) and 
        (c)FM-KPFM ($\df = -7.0$~Hz, $A=5$~nm$_\text{p-p}$, 
        $\Vac=1$~V, $f_\text{ac}=300$~Hz).  
        Scale bar is 1$\mu$m.
\label{fig:DKPFM_image}}
\end{figure*}

Figure~\ref{fig:DKPFM_image} shows CPD images 
of a patterned MoS$_2$ flake exfoliated on SiO$_2$/Si substrate 
taken by (a)$1\omega$D-KPFM, (b)$2\omega$D-KPFM and 
(c)FM-KPFM techniques taken with the same tip.
In $1\omega$ ($2\omega$) D-KPFM imaging, a sinusoidally oscillating voltage 
with $\wel=\wm$ ($\wel=2\wm$), $\Vac=50$~mV ($\Vac=1$~V) 
and $\phi=22^\circ$ ($\phi=18^\circ$)
coherent with the tip oscillation was applied to the sample.
In FM-KPFM imaging, a sinusoidally oscillating voltage 
with $\wel = 300$~Hz and $\Vac=1$~V
was applied to the sample.
The detection bandwidth of the lock-in was 100~Hz.
The scanning time for all the images were 1~s/line.
A stripe pattern with 2~$\mu$m pitch was created 
by reactive ion etching on the MoS$_2$ flake.
The topography images show an unetched terrace located 
between the etched regions (Image available in Supplemental Information.). 
The height of the unetched terrace is approximately 20~nm 
with respect to the etched regions.
The bands shown in the middle of the CPD images corresponds to the unetched
terrace and show a clear fractal-like pattern, 
which can be ascribed to the residue of the etch resist (PMMA).
All three CPD images show apparently a very similar pattern on the terrace.
However, a close inspection of the line profile of each image shows
that $2\omega$D-KPFM and FM-KPFM provide a very similar potential profile
with almost the same contrast
while $1\omega$D-KPFM shows a potential contrast about two times smaller
than that of $2\omega$D-KPFM and FM-KPFM.
The close similarity between $2\omega$D-KPFM and FM-KPFM originates from 
the fact that $\Fout$ (Eq.~\ref{eq:Fquad}) is proportional to 
the force gradient as we have discussed above.
Slightly larger potential contrast in $2\omega$D-KPFM compared with FM-KPFM
is due to the faster feedback response of $2\omega$D-KPFM 
by virtue of the absence of low frequency modulation \cite{Miyahara2015},
which is a clear advantage of both D-KPFM techniques over FM-KPFM.
The bandwidth of FM-KPFM is limited by detection bandwidth
of the lock-in amplifier that needs to be lower than the frequency of 
the ac bias voltage.
The lower contrast of $1\omega$D-KPFM is ascribed to its sensitivity to 
electrostatic force rather than electrostatic force gradient 
which results in larger spatial average due to the stray capacitance
including the body of the tip and the cantilever 
\cite{Hochwitz1996,Jacobs98, Glatzel2003a, Zerweck05, Nomura2007}.

In spite of lower potential contrast, $1\omega$D-KPFM 
has an advantage that it requires much smaller $\Vac= 50~$mV 
compared with 1~V required in $2\omega$D-KPFM and FM-KPFM.
This advantage is important for such samples as semiconductor 
where the influence of the large $\Vac$ can be very important 
due to band-bending effects.
As it is easy to switch between $1\omega$ and $2\omega$D-KPFM modes, 
$1\omega$D-KPFM can be used for qualitative measurement
with less electrical disturbance 
while $2\omega$D-KPFM can be used to obtain more accurate CPD contrast
on the same sample location.
Another advantage of $2\omega$D-KPFM is that it is free from the capacitive
crosstalk to the piezoelectric element and photodiode signal line 
\cite{Diesinger2010}
as the frequency of the ac voltage itself does not match any resonance 
of the cantilever.

In conclusion, we report a new technique that
enables force-gradient sensitive Kelvin probe force
microscopy using the dissipation signal of FM-AFM for dc voltage feedback.
It features the simpler implementation and faster scanning
as it requires no low frequency modulation.
The dissipation is caused by the oscillating electrostatic force
that is coherent with the tip oscillation,
which is induced by applying a sinusoidally oscillating ac voltage
with the frequency two times that of the tip oscillation frequency.
We theoretically analyzed the effect of the applied ac voltage
and show that the induced dissipation is sensitive to 
electrostatic force gradient rather than electrostatic force.
The experiments confirmed the theoretical analysis 
and demonstrated that $2\omega$D-KPFM provides essentially the same
result obtained by FM-KPFM.
The combination of $1\omega$ and $2\omega$D-KPFM techniques
will be a versatile tool to study the samples
whose electrical properties are sensitive to the external 
electric field.

The authors would like to thank Dr.~Omid Salehzadeh Einabad 
and Prof.~Zetian Mi at McGill University for 
providing the MoS$_2$ sample.
This work was partly supported by the Natural Science and Engineering
Research Council (NSERC), 
le Fonds Qu\'eb\'ecois de Recherche sur la Nature et
les Technologies (FQRNT).

\section*{References}


\begin{thebibliography}{19}
\bibitem [{not()}]{note:HF2LI}%
  \BibitemOpen
  \href@noop {} {}\bibinfo {note} {For convenience, HF2LI lock-in amplifier
  (Zurich Instruments) in external reference mode was used as a frequency
  doubler as well as phase shifter.}\BibitemShut {Stop}%
\end{thebibliography}
\end{document}


\title{Supplemental information for\\
''Force-gradient sensitive Kelvin probe force microscopy 
by dissipative electrostatic force modulation''}

\author{Yoichi Miyahara}%
\email[Corresponding author: ]{yoichi.miyahara@mcgill.ca}
\author{Peter Grutter}%
\affiliation{Department of Physics, Faculty of Science, McGill University, Montreal, Quebec, Canada H3A 2T8}

\date{\today}

\maketitle
\section{Derivation of Eq.~2 and 3}
In the absence of localized charges, 
the electrostatic force between a conducting tip and sample under dc and ac bias voltage
is described as follows:
\begin{eqnarray}
      \Felec (t) & = & \frac{1}{2}\dcdz \{\Vbias +\Vac \cos (\wel t + \phi) - \Vcpd\}^2 \\
      & = & \Fdc + \Fw + \Fww \nonumber \\
       \Fdc & = & \frac{1}{2} \dcdz \left[(\Vbias - \Vcpd)^2 + \frac{\Vac ^2}{2}\right] \label{Fdc} 
       = \alpha \left(\Vdc ^2 + \frac{\Vac^2}{2} \right)\\ 
       \Fw &= &  \dcdz (\Vbias - \Vcpd)\Vac \cos (\wel t + \phi) 
       = 2\alpha \Vdc \Vac \cos (\wel t + \phi)  \label{Fw} \\ 
       \Fww & = & \frac{1}{4} \dcdz \Vac ^2 \cos (2(\wel t + \phi))  \label{F2w} 
       = \frac{1}{2}\alpha \Vac^2 \cos (2(\wel t + \phi))
\end{eqnarray}
where
\[ 
  \label{eq:alpha}
         \alpha \equiv \frac{1}{2}\dcdz \nonumber, \,\,\,  \Vdc \equiv \Vbias - \Vcpd
\]

\begin{tabular}{l l l}
  $C$ & : & tip-sample capacitance\\
  $\Vbias$ & : &applied dc bias voltage\\
  $\Vcpd$ & : & contact potential difference\\
  $\Vac$ & : & amplitude of applied ac voltage\\
  $\wel$ & : & angular frequency of applied ac voltage\\
  $\phi$ & : & phase of applied ac voltage with respect to tip oscillation\\
  $z$ & : & position of the tip measured from the sample surface
\end{tabular}
\vspace*{5mm}

In general, $\alpha \equiv \frac{1}{2}\dcdz(z(t))$ is 
also a function of time through time-dependent $z(t)=A\cos(\wm t)$
with $\wm$ and $A$ being its oscillation angular frequency and amplitude,
respectively.

By expanding $\alpha$ around the mean position $z_0$,
and taking the first order,

\begin{equation}
  \alpha (z) \approx \alpha_0 + \alpha ' (z-z_0) = \alpha_0 + \alpha ' A \cos(\wm t)
\end{equation}

Substituting $\alpha$ into eq.\ref{Fdc} yields 
\begin{eqnarray}
       \Fdc(t)  &= & \{\alpha_0 + \alpha' A \cos(\wm t)\} \left(\Vdc ^2 + \frac{\Vac^2}{2}\right) \nonumber \\
       &=& \alpha_0 \left(\Vdc ^2 + \frac{\Vac^2}{2}\right)  
       + \alpha' A\left(\Vdc ^2 + \frac{\Vac^2}{2}\right) \cos (\wm t)
      \label{Fdc2} 
\end{eqnarray}

The first term in the above formula yields the static deflection of the cantilever 
wheares the second one is an alternating force whose frequency is $\wm/2\pi$.
As this force is in phase with $z(t)$, it will lead to the shift in the resonance frequency.
\\\\
Next, we look at $\Fw$ term (eq.~\ref{Fw}).
\begin{eqnarray*}
       \Fw(t) &= & 2 \{\alpha_0 + \alpha' A \cos(\wm t)\} \Vdc\Vac \cos (\wel t + \phi) \\
       & = & 2\alpha_0 \Vdc\Vac \cos(\wel t + \phi) 
       + 2\alpha' A\Vdc\Vac \cos(\wel t + \phi) \cos(\wm t) \nonumber \\
       & = & 2\alpha_0 \Vdc\Vac \cos(\wel t + \phi) 
       + 2\alpha' A\Vdc\Vac \frac{1}{2}[\cos\{(\wel + \wm) t + \phi)\} + \cos\{(\wel - \wm)t +\phi\}] \nonumber \\
       & = & 2\alpha_0 \Vdc\Vac \{\cos(\wel t)\cos(\phi)- \sin(\wel t)\sin(\phi) \} \nonumber \\
       & & + \alpha' A\Vdc\Vac [\cos\{(\wel + \wm) t + \phi)\} + \cos\{(\wel - \wm)t +\phi\}]       \label{Fw2} 
\end{eqnarray*}

When $\wel =2 \wm$, 
\begin{eqnarray}
       \Fw(t) 
       & = & 2\alpha_0 \Vdc\Vac \{\cos(2\wm t)\cos(\phi)- \sin(2\wm t)\sin(\phi) \} \nonumber \\
       & & + \alpha' A\Vdc\Vac [\cos\{(3\wm) t + \phi)\} + \cos(\wm t +\phi)]       \label{Fw2} 
\end{eqnarray}
The only component that contributes to the frequency shift and dissipation is the last term
in Eq.~\ref{Fw2}.

Looking at $\Fww$ term in the same way,
\begin{eqnarray}
       \Fww(t) &=& \frac{1}{2} \{\alpha_0 +  \alpha' A \cos(\wm t)\} \Vac ^2 \cos (2(\wel t + \phi)) \nonumber \\ 
       & = & \frac{1}{2}  [\alpha_0 \Vac ^2 \cos (2(\wel t + \phi)) 
       + \frac{1}{2} \alpha' A \Vac ^2 \{ \cos \{(2\wel - \wm)t + 2\phi)\} 
       +  \cos \{(2\wel + \wm) t + 2\phi)\}]  \nonumber \\
       & = & \frac{1}{2}  \alpha_0 \Vac ^2 \cos (2(\wel t + \phi)) 
       + \frac{1}{4} \alpha' A \Vac ^2 \{ \cos ((2\wel - \wm)t + 2\phi)
       +  \cos ((2\wel + \wm) t + 2\phi)\} \nonumber \\
       & = & \frac{1}{2}  \alpha_0 \Vac ^2 \cos (4\wm t + 2\phi) 
       + \frac{1}{4} \alpha' A \Vac ^2 \{ \cos (3\wm t + 2\phi)
       +  \cos (5\wm t + 2\phi)\}   \label{eq:F2w2}
\end{eqnarray}

 No term in Eq.~\ref{eq:F2w2} contributes to the frequency shift and dissipation.

 Gathering the terms with $\wm$ from Eq.~\ref{Fdc2} and \ref{eq:F2w2}, 
we obtain the following result shown in Eq~2 and 3 in the body text.

 \begin{eqnarray*}
  \Fwm(t) & = &   \alpha' A \left\{ \left(\Vdc ^2 + \frac{\Vac^2}{2}\right) \cos (\wm t)
      +  \Vdc \Vac  \cos (\wm t + \phi)\right\}\\
       & = & \alpha' A \left\{\Vdc ^2 +  \cos(\phi) \Vdc\Vac +\frac{\Vac^2}{2} \right\}\cos(\wm t) \\ 
  & &    -\alpha' A \Vdc \Vac  \sin(\phi) \sin(\wm t)\\
    & = & \Fin \cos (\wm t) + \Fout \sin(\wm t)
\end{eqnarray*}
where
\begin{equation}
  \Fin =  \alpha'A \left(\Vdc + \frac{\Vac}{2} \cos\phi   \right)^2
  + \frac{\alpha'A }{2}\left(1 - \frac{\cos^2\phi}{2} \right) \Vac^2
  \label{eq:Fin}
\end{equation}

\begin{equation}
  \Fout = - \alpha'A  \Vdc\Vac  \sin\phi 
    \label{eq:Fquad}
\end{equation}

\section{Fitting result of frequency shift offset as a function of phase}
\begin{figure}[h]
  \centering
  \includegraphics[width=80mm]{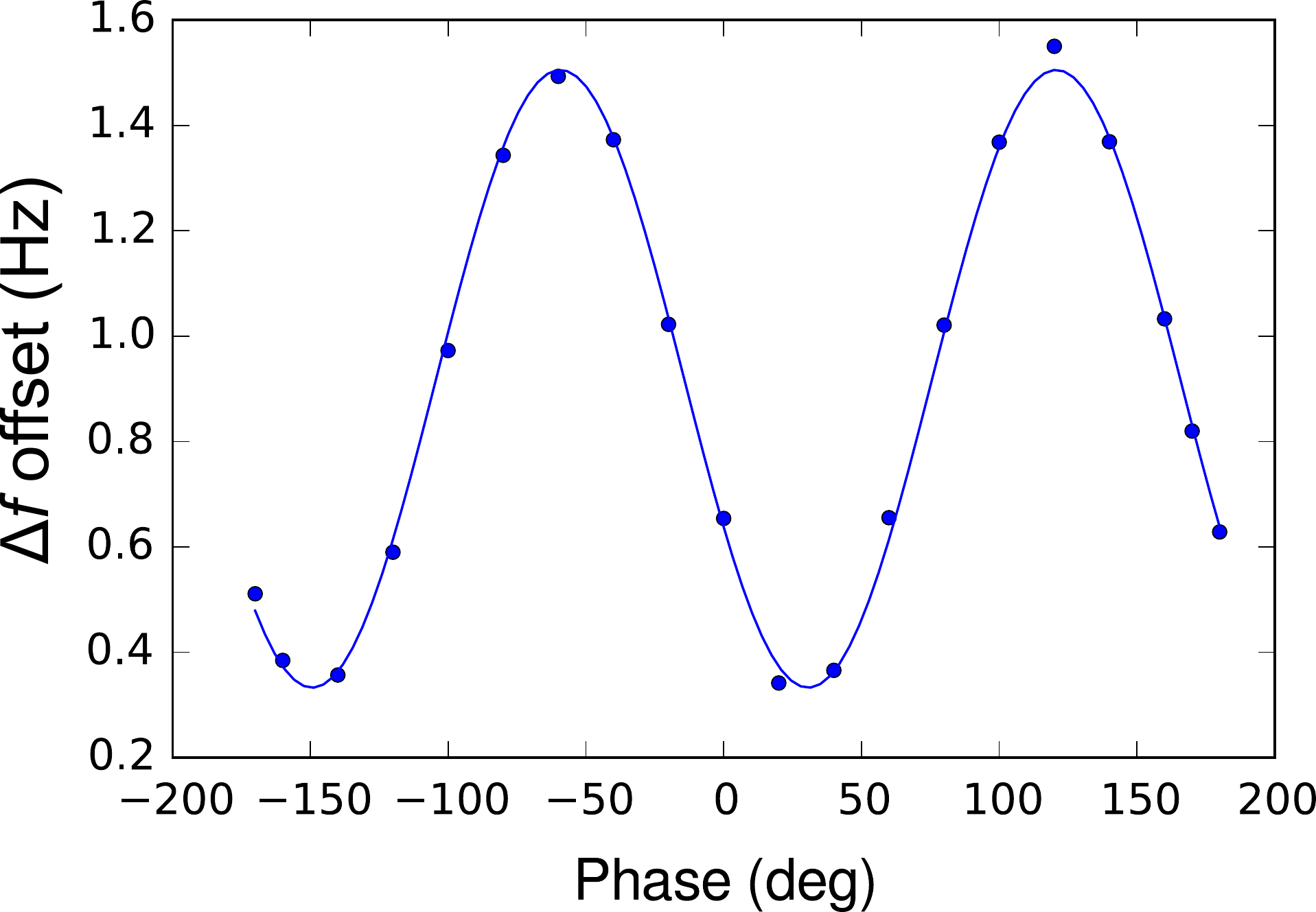}
  \caption{Frequency shift offset as a function of phase. 
    Solid line is the fitted curve with the second term of Eq.~2 in the body text.
\label{fig:DKPFM_image}}
\end{figure}

\section{Detail of the sample}
\subsection{Optical micrograph of MoS$_2$ flakes}
\begin{figure}[h]
  \centering
  \includegraphics[width=80mm]{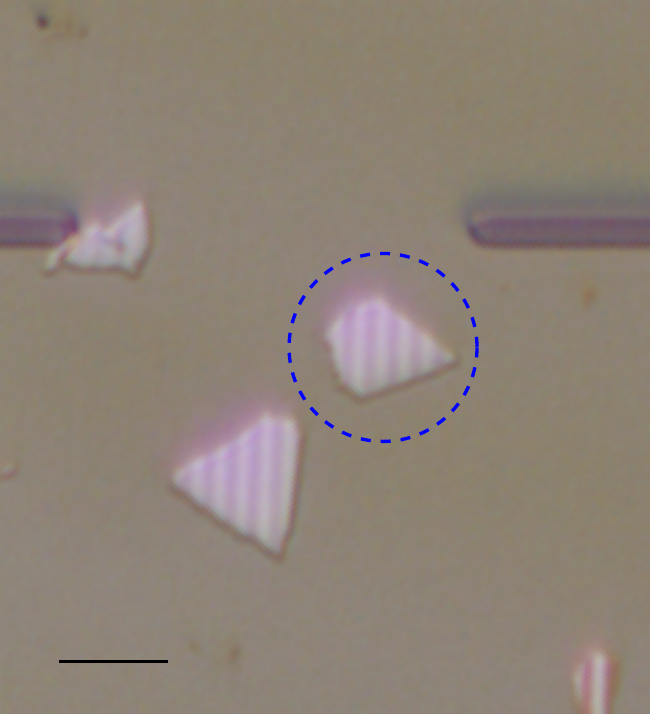}
  \caption{Optical micrograph of the MoS$_2$ flakes exfoliated on SiO$_2$/Si substrate. 
The flake in the circle is the one imaged by KPFM. Scale bar is 20~$\mu$m.
\label{fig:DKPFM_image}}
\end{figure}

\subsection{AFM topography images of MoS$_2$ flake}
\begin{figure}[h]
  \centering
  \includegraphics[width=170mm]{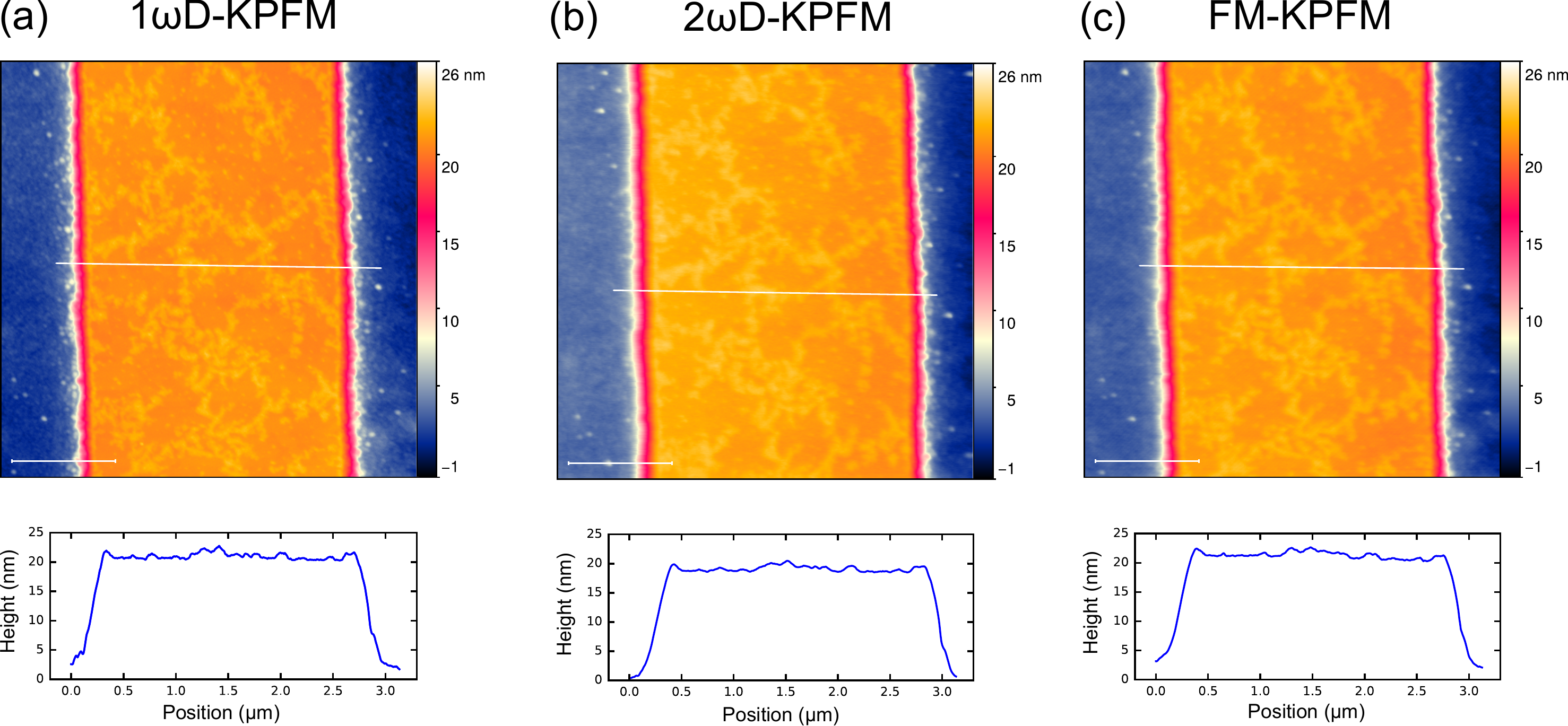}
  \caption{Topography images of patterned MoS$_2$ on SiO$_2$/Si substrate 
        simultaneously taken with each corresponding KPFM image shown in 
        Fig.~4 in the body text. 
        (a) D-KPFM ($1\omega$) ($\df = -10.3$~Hz, $A=5$~nm$_\text{p-p}$, 
        $\Vac=50$~mV) and 
        (b) D-KPFM ($2\omega$) ($\df = -8.3$~Hz, $A=5$~nm$_\text{p-p}$, 
        $\Vac=1$~V) and 
        (b) FM-KPFM ($\df = -7.0$~Hz, $A=5$~nm$_\text{p-p}$, 
        $\Vac=1$~V). 
        The line profiles are shown for the same location on the sample.
\label{fig:DKPFM_image}}
\end{figure}

\bibliography {../library}